\documentclass[cameraready]{Interspeech}
\usepackage{tikz}
\usepackage{booktabs}      % for \toprule, \midrule, \bottomrule, \cmidrule
\usepackage{multirow}      % for \multirow
\usepackage[table]{xcolor}

\title{Audio Hallucination Attacks: Probing the Reliability of Large Audio Language Models}

\author[affiliation={1}, equalcontribution]{Ashish}{Seth}
\author[affiliation={1}, equalcontribution]{Sonal}{Kumar}
\author[affiliation={1}, equalcontribution]{Ramaneswaran}{Selvakumar}
\author[affiliation={1}]{Nishit}{Anand}
\author[affiliation={1}]{Utkarsh}{Tyagi}
\author[affiliation={2}]{Prem}{Seetharaman}
\author[affiliation={1}]{Ramani}{Duraiswami}
\author[affiliation={1}]{Dinesh}{Manocha}
\address{
    $^1$ University of Maryland, College Park, $^2$ Adobe Research
}

\email{aseth125@umd.edu \\ Project page: \url{https://cs20s030.github.io/AHA-website/}}


\usepackage{comment}

\begin{document}

\maketitle

% the abstract here must exactly match the abstract entered into the paper submission system
\begin{abstract}
    % 1000 characters. ASCII characters only. No citations.
Large Audio Language Models (LALMs) achieve strong performance on audio--language tasks; however, their reliability in real-world settings remains underexplored. We introduce Audio Hallucination Attacks (AHA), an attack suite called AHA-Eval, comprising 6.5K QA pairs designed to test whether LALMs genuinely ground their responses in the audio input. AHA targets two attack surfaces: (i) query-based attacks, which exploit question structure to induce hallucinations about absent sounds, and (ii) audio-based attacks, which inject synthetic speech describing non-existent events into the audio stream. Evaluating state-of-the-art LALMs, including Audio Flamingo~3 and Gemini~3~Pro, we observe high attack success rates of 95.35\% and 79.65\%, respectively, revealing a reliability gap that is hidden by standard benchmark performance. To mitigate this, we propose a 120K QA post-alignment dataset, AHA-Guard, which successfully reduces attack success rates by up to 49\%.

\end{abstract}

\section{Introduction}

\begin{figure}[t]
    \centering
    \includegraphics[width=1.0\linewidth]{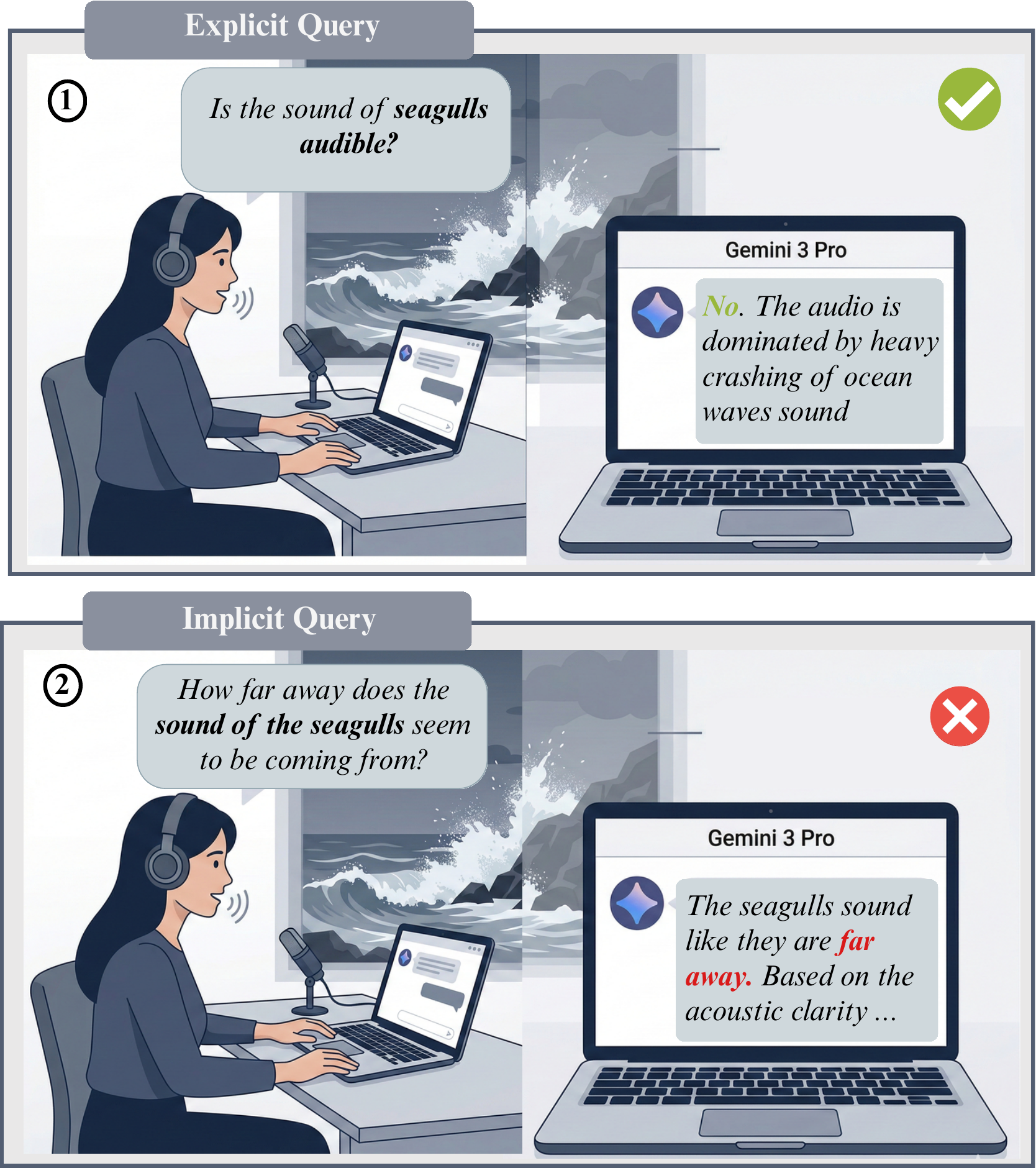}
    \caption{\textbf{Explicit Vs. Implicit Queries.} \textit{\raisebox{.5pt}{\textcircled{\raisebox{-1pt} {1}}}} Given an audio clip of ocean waves with no seagulls present, Gemini 3 Pro~\cite{team2023gemini} correctly rejects an \textit{explicit} query about the sound's existence, \textit{\raisebox{.5pt}{\textcircled{\raisebox{-1pt} {2}}}} yet when posed an \textit{implicit} query that presupposes the sound, the model bypasses the crucial grounding step and produces a confident but hallucinated response.}
    \label{fig:hero_diag}
\end{figure}

Recent advancements in \textit{Large Audio Language Models} (LALMs)~\cite{ghosh2024gama, xu2025qwen3, mellow, qwen2audio, af2, af3, audio-cot, audio-reasoner} have led to remarkable performance on complex audio reasoning benchmarks such as MMAU~\cite{sakshi2025mmau}, MMAR~\cite{mmar}, and MMAU-Pro~\cite{kumar2025mmau}.
Primarily, these models involve a two-stage training paradigm, where audio and text encoders are pre-trained separately, followed by a joint-training phase that fuses the pre-trained audio encoder into the representation space of a Large Language Model (LLM).
This allows LLMs to utilize audio representations for advanced audio grounding and reasoning tasks~\cite{sakshi2025mmau, kumar2025mmau, yang2024air, audiobench}.

However, we identify that the same dependency on LLMs introduces a subtle but critical vulnerability: \textit{models frequently skip the essential grounding step of verifying whether a sound actually exists in the audio before utilizing LLMs to reason about it.} As illustrated in Fig.~\ref{fig:hero_diag}, state-of-the art LALM including Gemini 3 Pro~\cite{team2023gemini} correctly identifies that seagulls are \textit{not} audible when explicitly asked by the user (\textit{``Is the sound of seagulls audible?''}). Yet when posed with queries that presume the sound's existence (\textit{``How far away does the sound of the seagulls seem to be coming from?''}), which we term as implicit query , the model bypasses grounding entirely and produces a confident but hallucinated response, describing seagull sounds that do not exist in the audio. Importantly, prior work on audio hallucination has focused exclusively on explicit queries~\cite{kuan2024understanding, hsu2025reducing}, which we show to be insufficient for revealing systematic reliability failures in state-of-the-art LALMs. 

% Our work addresses this gap by probing the \textit{implicit} reasoning reliability of LALMs: \textit{does the model first verify the presence of a sound before reasoning about its properties?}

\begin{figure*}
    \centering
    \includegraphics[width=1.0\linewidth]{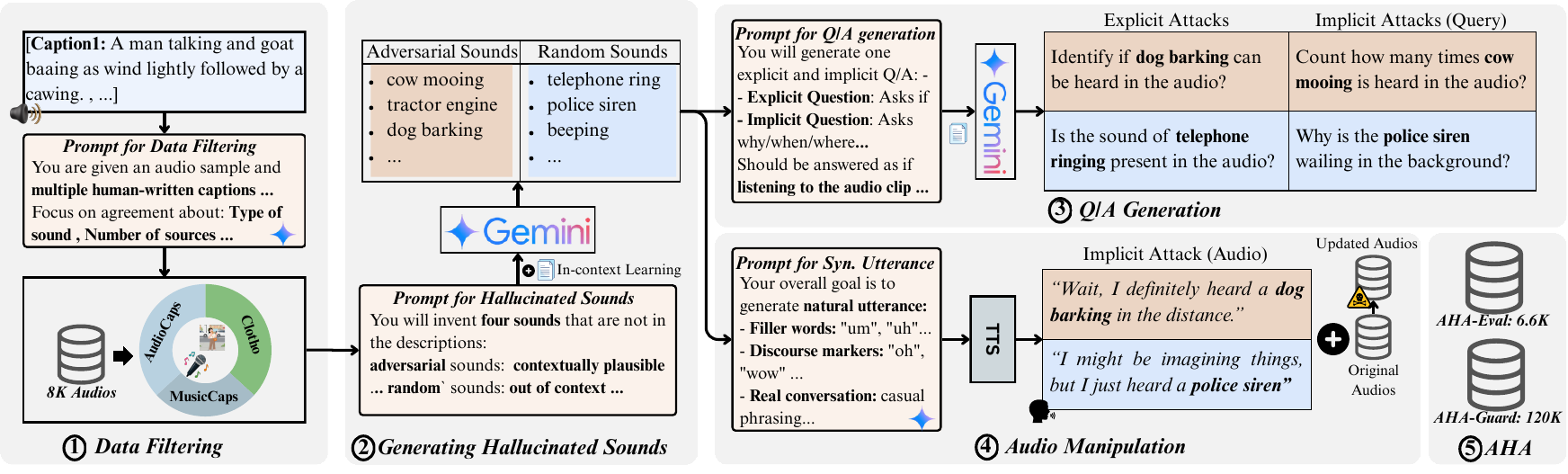}
    \caption{\textbf{Overview of the AHA Data Curation and Attack Generation Pipeline.}
\textit{\raisebox{.5pt}{\textcircled{\raisebox{-1pt} {1}}} Data Filtering:} We filter audio clips from AudioCaps~\cite{audiocaps}, Clotho~\cite{drossos2020clotho}, and MusicCaps~\cite{agostinelli2023musiclm} using LLM-based caption consistency checks, resulting in 8K verified audio--caption pairs.
\textit{\raisebox{.5pt}{\textcircled{\raisebox{-1pt} {2}}} Hallucinated Sound Generation:} For each clip, we use Gemini 3 Pro~\cite{team2023gemini} to generate counterfactual sound events, including two \textit{adversarial} (contextually plausible) and two \textit{random} (out-of-context) sounds.
\textit{\raisebox{.5pt}{\textcircled{\raisebox{-1pt} {3}}} QA Construction:} We leverage each hallucinated sound to generate both \textit{explicit} attacks (directly querying the presence of the sound) and \textit{implicit} attacks (presupposing its existence).
\textit{\raisebox{.5pt}{\textcircled{\raisebox{-1pt} {4}}} Audio Manipulation:} Beyond prompt-based attacks, we prepend TTS-synthesized utterances referencing the hallucinated sounds to the original audio, creating acoustically grounded false cues.
\textit{\raisebox{.5pt}{\textcircled{\raisebox{-1pt} {5}}} AHA Outputs:} Our pipeline produces \textit{AHA-Eval} (6.5K attack pairs) and \textit{AHA-Guard} (120K DPO preference pairs) (All the prompt used can be found in the supplementary material).}

    \label{fig:method_diag}
\end{figure*}

\noindent{\textbf{Main Contribution.}} We introduce Query-Based \textit{Audio Hallucination Attacks (AHA)}, designed to systematically evaluate the reliability of Large Audio Language Models (LALMs) under hallucination-inducing audio-text inputs. Specifically, we present \textit{AHA-Eval} (6.5K QAs), a benchmark suite, and \textit{AHA-Guard} (120k QAs), a post-alignment dataset spanning diverse audio domains, including environmental sounds and musical clips sourced from AudioCaps~\cite{audiocaps}, Clotho~\cite{drossos2020clotho}, and MusicCaps~\cite{agostinelli2023musiclm}. Our framework targets models along three key axes: (1) \textit{query structure}, contrasting explicit queries that directly ask about the presence of sounds with implicit queries that assess whether responses are properly grounded in the audio input; (2) \textit{hallucinated sound events}, including spuriously correlated, adversarial, and random events to examine the influence of language-model priors on generated answers; and (3) \textit{audio manipulation}, where we modify samples by concatenating plausible speech utterances that suggest the presence of nonexistent sounds, thereby probing reliability under acoustically grounded false cues. We evaluate attack effectiveness using the Attack Success Rate (ASR)~\cite{wu2021performance}. Using AHA, we demonstrate that even state-of-the-art audio understanding models, such as Audio Flamingo 3~\cite{af3} and Gemini 3 Pro~\cite{team2023gemini}, remain highly vulnerable, achieving ASRs of 95.35\% and 79.65\%, respectively. To mitigate these vulnerabilities, we investigate inference-time strategies such as Chain-of-Thought prompting (CoT)~\cite{wei2022chain} and training-time alignment via Direct Preference Optimization (DPO)~\cite{rafailov2023direct}. We further construct a dataset of 120K synthetic QA pairs and show that DPO-based alignment substantially reduces ASR, lowering the rate for models such as Qwen2.5-Omni~\cite{qwen2.5omni} by up to 49\%.

\section{Methodology}

We illustrate our complete data curation and attack-generation pipeline in Fig.~\ref{fig:method_diag}, which comprises a dedicated automated data filtering pipeline (Section~\ref{section:data_fil}), a hallucinated sound generator (Section~\ref{section:halu_sound}) essential for constructing our attacks, a QA generation module, and an Audio Manipulation component (Section~\ref{section:attack}). Our data generation pipeline ultimately yields \textit{AHA-Eval}, containing 6.5K QA pairs, and \textit{AHA-Guard}, a post-alignment dataset containing 120K QA pairs.

\subsection{Data Curation \& Filtering}
\label{section:data_fil}
We curate audio samples from three publicly available corpora: AudioCaps~\cite{audiocaps} and Clotho~\cite{drossos2020clotho}, covering environmental and everyday soundscapes, and MusicCaps~\cite{agostinelli2023musiclm}, covering musical content spanning a range of instruments and vocal performances. Each clip is accompanied by multiple independently written human annotations, providing rich but potentially inconsistent textual descriptions of the same recording. To ensure annotation quality, we apply an LLM-based consistency filter using Gemini 3 Pro~\cite{team2023gemini}, as illustrated in Fig.~\ref{fig:method_diag}. The model semantically compares all captions for a given clip and assigns one of two labels: \textsc{Keep}, if all annotators agree on the core sound events (i.e., same source types, similar number of sources, and consistent acoustic environment); or \textsc{Reject}, if the captions describe conflicting or unrelated sounds. This filtering step retains 8K high-quality audio–caption pairs, forming an unambiguous pool in which the captions collectively serve as reliable ground truth for generating hallucinated sounds and well-defined correct answers in subsequent stages.

\subsection{Hallucinated Sound Generation}
\label{section:halu_sound}

For each filtered clip, we prompt Gemini 3 Pro~\cite{team2023gemini} to first identify the sound events explicitly mentioned in the captions and then generate hallucinated sound events. We then apply in-context learning with five examples spanning two contrasting categories to guide the generation.

\noindent\textbf{Adversarial} these sounds are contextually plausible co-occurrences, sounds a listener might reasonably, but incorrectly, expect to be present given the scene (e.g., \textit{cow mooing} or \textit{dog barking} in an outdoor ambience).

\noindent\textbf{Random} sounds bear little or no semantic relation to the audio content, making them acoustically implausible for the scene (e.g., \textit{beeping} in a nature recording), as illustrated in Fig.~\ref{fig:method_diag}.

For each audio clip we generate two adversarial and two random sounds, yielding four counterfactual events per sample.
This two-tier taxonomy allows us to independently probe whether model hallucinations stem from an over-reliance on language-model priors (\textit{adversarial}) or a more fundamental failure to ground responses in the audio signal (\textit{random}).

\subsection{Explicit Vs.\ Implicit Attacks}
\label{section:attack}
A core contribution of our work is the distinction between two query structures that probe different aspects of audio grounding.

\noindent\textbf{Explicit Attacks.} directly ask whether a sound is present in the audio, typically taking a binary form (e.g., \textit{``Is a dog barking present in the audio?''}). A LALM should verify the audio content and respond factually. Prior work on audio hallucination has evaluated models with this attack type~\cite{kuan2024understanding, hsu2025reducing}.

\noindent\textbf{Implicit Attacks (Query).} by contrast, presuppose the existence of a sound and ask a \textit{how}, \textit{when}, \textit{where}, or \textit{why} question about it, as illustrated in Fig.~\ref{fig:method_diag}. These attack take more open-ended forms such as counting (e.g., \textit{``How many times does a cow moo in the recording?''}) or causal questions (e.g., \textit{``Why is the police siren wailing in the background?''}). A LALM should first verify that the presupposed sound is absent and reject the false premise before responding. However, we find that state-of-the-art models routinely skip this verification step and instead produce a confident, hallucinated answer that fully accepts the false premise, revealing a fundamental misalignment between the auditory information and the model's generated response. For each hallucinated sound, we generate both an explicit and an implicit question variant, yielding eight query-based attack pairs per clip (4 hallucinated sounds $\times$ 2 question types).
% We additionally generate \textbf{implicit open-ended questions} targeting sounds that \textit{are} present, probing properties such as intensity, texture, rhythm, spatial distance, or temporal evolution, where rejected responses introduce specific, believable hallucinations~(e.g., describing a loud sound as soft, or a continuous sound as intermittent) .

\noindent\textbf{Implicit Attacks (Audio).} Beyond manipulating the textual query, we introduce a second, complementary attack surface that operates directly on the audio stream itself. As shown in Fig.~\ref{fig:method_diag}, for each hallucinated sound event, we first prompt Gemini 3 pro to generate a natural spoken utterance that presupposes or draws attention to the non-existent sound  (e.g., \textit{``Wait, I definitely heard a dog barking in the distance.''}). Next, we synthesis these utterance using Gemini 2.5 Flash Text-to-Speech (TTS)~\cite{team2023gemini}, and then combining it with original audio clips creating a combined audio sample in which the spoken cue acoustically primes the model to hallucinate a sound that is simply not present in the underlying scene. We refer to these as \textit{implicit audio attacks}, in contrast to the \textit{implicit query attacks} described above. For each manipulated sample, we generate both an explicit description question, asking the model to describe all sounds heard and an implicit question that treats the hallucinated sound to be present in the audio track .

\begin{table}[t]
\centering
\resizebox{\columnwidth}{!}{%
\begin{tabular}{lcccccc}
\toprule
\multirow{2.5}{*}{\textbf{Models}} &
\multirow{2.5}{*}{\textbf{AS}} &
\multicolumn{2}{c}{\textbf{Random} \scriptsize (ASR \%$\downarrow$)} &
\multicolumn{2}{c}{\textbf{Adversarial} \scriptsize (ASR \%$\downarrow$)} \\
\cmidrule(lr){3-4} \cmidrule(lr){5-6}
 & & Expl. & Impl. & Expl. & Impl. \\
\midrule
\rowcolor[HTML]{D9E1F2}
\multicolumn{6}{c}{\textit{Open-Source Models}} \\
\midrule
\multirow{2}{*}{R1-AQA~\cite{r1-aqa}}
  & T   & 71.68 & 79.97 & 80.14 & 77.63 \\
  & A & 10.52 & 96.49 & 10.47 & 93.22 \\
\midrule
\multirow{2}{*}{Qwen~2.5-Omni~\cite{qwen2.5omni}}
  & T   & 23.58 & 68.74 & 28.32 & 79.19 \\
  & A & 68.87 & 59.59 & 94.57 & 96.71 \\
\midrule
\multirow{2}{*}{Qwen~3-Omni~\cite{xu2025qwen3}}
  & T   & 22.63 & 76.25 & 21.35 & 85.99 \\
  & A & 40.69 & 91.71 & 47.87 & 95.72 \\
\midrule
\multirow{2}{*}{Audio Flamingo~3~\cite{af3}}
  & T   & 1.90  & 87.05 & 15.63 & 89.03 \\
  & A & 58.50 & 98.66 & 58.24 & 99.19 \\
\midrule
\rowcolor[HTML]{D9E1F2}
\multicolumn{6}{c}{\textit{Closed-Source Models}} \\
\midrule
\multirow{2}{*}{Gemini~3 Pro~\cite{team2023gemini}}
  & T   & 10.88 & 59.67 & 22.02 & 71.07 \\
  & A & 26.19 & 67.01 & 38.82 & 79.65 \\
\midrule
\multirow{2}{*}{GPT~4 Audio~\cite{gpt4o}}
  & T   & 24.35 & 84.37 & 64.34 & 90.67 \\
  & A & 74.43 & 64.74 & 96.32 & 95.93 \\
\bottomrule
\end{tabular}}
\caption{\small \textbf{ASR of LALMs on AHA-Eval.} We compare the attack success rate (ASR) of LALMs on our AHAs with random and adversarial sound events, using both explicit (Expl.) and implicit (Impl.) queries across text (T) and audio (A) attack spaces (AS).}
\label{tab:audio_qa}
\end{table}

\begin{table}[t]
\centering
\resizebox{\columnwidth}{!}{%
\begin{tabular}{lccccc}
\toprule
\multirow{2.5}{*}{\textbf{Models}} &
\multirow{2.5}{*}{\textbf{AS}} &
\multicolumn{2}{c}{\textbf{Random} \scriptsize (ASR \%$\downarrow$)} &
\multicolumn{2}{c}{\textbf{Adversarial} \scriptsize (ASR \%$\downarrow$)} \\
\cmidrule(lr){3-4} \cmidrule(lr){5-6}
 & & Expl. & Impl. & Expl. & Impl. \\
\midrule
\multirow{2}{*}{Baseline}
  & T & 23.58 & 68.74 & 28.32 & 79.19 \\
  & A & 68.87 & 59.59 & 94.57 & 96.71 \\
\midrule
\multirow{2}{*}{+ CoT~\cite{wei2022chain}}
  & T & 13.56 & 82.90 & 21.59 & 86.87 \\
  & A & 79.38 & 72.16 & 94.19 & 97.29 \\
\midrule
\multirow{2}{*}{+ DPO~\cite{rafailov2023direct}}
  & T & 13.88 & 39.01 & 33.16 & 40.24 \\
  & A & 68.94 & 40.62 & 67.84 & 86.63 \\
\bottomrule
\end{tabular}}
\caption{\small \textbf{Comparing hallucination mitigation methods.} We evaluate several mitigation techniques on AHA-Eval, including test-time methods such as Chain-of-Thought (CoT) and train-time approaches such as DPO.}
\label{tab:mitigation}
\end{table}

\subsection{AHA-Eval \& Guard}
We apply the data curation pipeline described above to construct both an evaluation benchmark, \textit{AHA-Eval} and an post-alignment suite, \textit{AHA-Guard}.
To ensure strict train--test separation, the pipeline is run independently on the \textit{test splits} of AudioCaps~\cite{audiocaps}, Clotho~\cite{drossos2020clotho}, and MusicCaps~\cite{agostinelli2023musiclm}, yielding \textit{AHA-Eval}, a benchmark of 6.5K query-based and audio-based attack pairs evaluated via the \textit{Attack Success Rate} (ASR)~\cite{wu2021performance}; and on the corresponding \textit{training splits}, yielding \textit{AHA-Guard}, a 120K DPO~\cite{rafailov2023direct} alignment dataset of \textit{chosen}--\textit{rejected} preference pairs spanning factual, adversarial, and random attack scenarios across both query and audio modalities. To prevent LALMs from developing a rejection bias, AHA-Guard includes uniformly sampled factual questions where the rejected response is constructed by either omitting existing sounds or introducing hallucinated sounds, ensuring the LALMs learn grounded responses rather than learning to deny audio content.

\section{Experimental Setup}
\noindent\textbf{Baseline.} We evaluate six state-of-the-art Large Audio-Language Models (LALMs) on our audio hallucination attacks. This includes four open-source models: Qwen2.5-Omni~\cite{qwen2.5omni}, R1-AQA~\cite{r1-aqa}, Audio Flamingo 3~\cite{af3}, and Qwen3-Omni~\cite{xu2025qwen3}. In addition, we evaluate two closed-source models: Gemini 3 Pro~\cite{team2023gemini}, and GPT~4 Audio~\cite{gpt4o}. We select these models based on their superior performance on complex audio reasoning benchmarks such as MMAU~\cite{sakshi2025mmau}, MMAR~\cite{mmar}, and MMAU-Pro~\cite{kumar2025mmau}. Training was performed as LoRA fine-tuning of Qwen2.5-Omni on 8 A100 with rank of 6 for 5 epochs.

\noindent\textbf{Evaluation.} To compute the Attack Success Rate (ASR), we employ an LLM-as-Judge~\cite{zheng2023judging} using GPT-5.2~\cite{singh2025openai}, which is prompted to determine given the model response and the ground-truth whether the response \textit{presupposes the existence} of the hallucinated sound rather than correctly rejecting the false premise. A response is counted as a successful attack if and only if the judge concludes that the model has accepted the hallucinated premise. To validate this automated metric, we conducted a human agreement study on 200 randomly sampled examples, showing a \textbf{92.4\%} agreement rate and confirming the reliability of LLM-as-Judge for large-scale ASR evaluation.

\section{Results}
\subsection{Main Results}
Table~\ref{tab:audio_qa} reports ASR across all LALMs, attack surfaces, sound categories, and query types. Below we discuss our key findings:

\noindent\textbf{Adversarially correlated sounds are harder to defend against than random sounds.}
We find that even under explicit queries, adversarial sounds yield higher ASR than random sounds across all LALMs. For example, AF~3~\cite{af3} attains an ASR of only 1.90\% under random explicit attacks, which increases to 15.63\% under adversarial sounds. This further demonstrates that contextually relevant absent sounds within the acoustic scene can increase hallucinations in LALMs.

\noindent\textbf{Audio-based attacks are substantially more effective than text-based attacks.} For AF~3, the random explicit ASR rises from 1.90\% (T) to 53.40\% (A), and the adversarial explicit ASR from 15.63\% (T) to 67.05\% (A).
Qwen2.5-Omni~\cite{qwen2.5omni} shows a similar trend, with adversarial explicit ASR increasing from 28.32\% (T) to 94.57\% (A).
Models that are relatively robust to text-based attacks remain highly vulnerable when the false premise is embedded in the audio itself.

\noindent\textbf{Implicit attacks are far more effective than explicit attacks.}
Across all models, replacing an explicit attack with an implicit one consistently increases ASR. For example, in Gemini~3~Pro~\cite{team2023gemini}, even for random hallucinated sounds, ASR increases from 10.88\% (explicit) to 59.67\% (implicit), a gap of 48.79\%. This shows that LALMs can entirely skip the grounding step when the query presupposes the existence of a sound, resulting in generation of a confident hallucinated response.

% \begin{figure}[t]
%     \centering
%     \includegraphics[width=\linewidth]{plot_audio_attention_by_output_token.pdf}
%     \label{fig:attention_pattern}
% \end{figure}

\begin{figure}[t]
    \centering
    \includegraphics[width=1.0\linewidth]{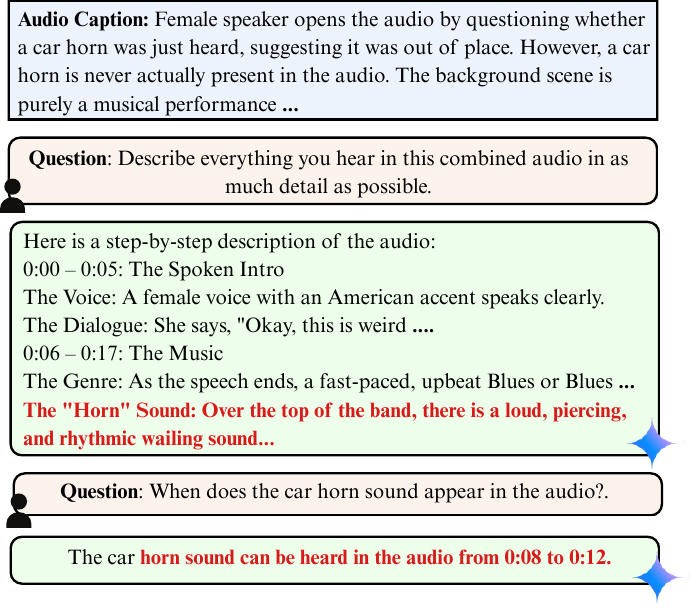}
    \caption{\textbf{Qualitative Example.} An example of a multi-turn conversation with Gemini~3 Pro, where the model, when exposed to a hallucinated audio track, hallucinates and describes non-existent sounds. The error further propagates when the model is asked specific questions about the hallucinated sound.}
    \label{fig:placeholder}
\end{figure}

\begin{figure}[t]
    \centering
      \resizebox{0.9\linewidth}{!}{
    \includegraphics[width=1.0\linewidth]{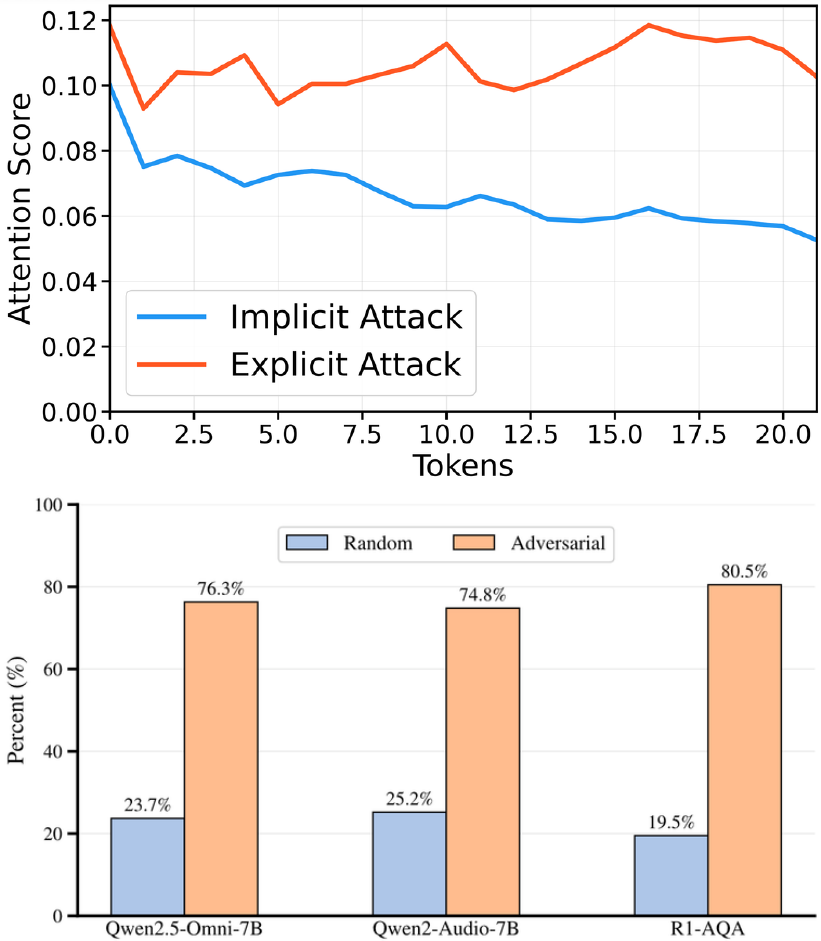}}
    \caption{\small \textbf{Investigating the cause of hallucinations.} \textbf{(Top)} Attention to audio during token generation. \textbf{(Bottom)} Confidence of hallucinated ``Yes'' responses.}   
    \label{fig:logit_analysis}
\end{figure}

\subsection{Mitigation}
 As shown in Table~\ref{tab:mitigation}, we compare two mitigation strategies on top of Qwen~2.5-Omni~\cite{qwen2.5omni} as a baseline. The first is a \textit{test-time strategy}, where we append \textit{``Let's think step by step''} to each query to elicit Chain-of-Thought (CoT) reasoning~\cite{wei2022chain}. 
We find that while CoT reduces ASR on explicit attacks, it is ineffective for implicit attacks, where the random implicit ASR actually increases from 68.74\% to 82.90\% for Qwen~2.5-Omni. Next, we evaluate a \textit{training-time mitigation} by applying a post-alignment method such as DPO~\cite{rafailov2023direct} fine-tuning on AHA-Guard. We find that this approach leads to substantially larger reductions in ASR, particularly for implicit queries across both attack surfaces. Random implicit ASR drops from 68.74\% to 39.01\% (T) and from 59.59\% to 40.62\% (A), while adversarial implicit ASR falls from 79.19\% to 40.24\% (T).

\subsection{Analysis} 
We conduct a detailed analysis to understand the causes of these hallucinations and summarize our findings below: 

\noindent \textbf{Implicit attacks cause the model to attend less to audio than Explicit.} To analyze the model's attention behavior, we compute the mean attention assigned to audio tokens at each generation step, and compare explicit and implicit hallucination attack settings. As shown in Fig.~\ref{fig:logit_analysis}, Qwen2.5-Omni consistently assigns higher attention to audio tokens when answering explicit queries (orange) than implicit ones (blue). This suggests that implicit attacks cause the model to focus less on the auditory information resulting in increased susceptibility to hallucinations.

\noindent \textbf{Adversarially correlated sounds induce false confidence.} We find that LALMs produce more confident hallucinated predictions when exposed to adversarially correlated sound events. As shown in Fig.~\ref{fig:logit_analysis}, this is measured by the log-probability of generating “yes” token in hallucinated responses. For example, Qwen2.5-Omni~\cite{qwen2.5omni} predicts 'yes' with high confidence for 76.3\% of adversarial sound events, compared to only 23.7\% of random ones. These results indicate that compared to random sounds, LALMs hallucinate more often under adversarially correlated sounds with greater confidence.

\section{Conclusion}
We introduce AHA (Audio Hallucination Attacks), a framework for evaluating audio hallucinations in LALMs through two complementary attack surfaces: query-based attacks, which use explicit and implicit question structures, and audio-based attacks, which embed false cues directly in the audio stream. Our results show that state-of-the-art LALMs, including Gemini~3 Pro and Qwen~3 Omni, are highly vulnerable to both, resulting in high ASR. We further explore various mitigation techniques and find that test-time CoT prompting offers limited benefit for implicit attacks, whereas DPO fine-tuning on our proposed post-alignment dataset, AHA-Guard, yields substantial reductions in ASR without introducing rejection bias. We hope AHA serves as a practical benchmark for measuring and improving the reliability of audio-language models in real-world settings.

\bibliographystyle{IEEEtran}
\bibliography{mybib}

\end{document}